\documentclass{iopjournal}

\usepackage{hyperref}
\usepackage{graphicx}
\usepackage{dcolumn}
\usepackage{bm}
\usepackage{physics}
\usepackage{afterpage}
\usepackage{orcidlink}
\usepackage{cite}

\def\eps{{\varepsilon}}

\begin{document}

\articletype{Paper} 

\title{From phase synchronization to waveform proportionality in a population of R\"ossler oscillators driven by an external pacemaker}

\author{Yuzuru Mitsui$^{1, 2, *}$\orcidlink{0009-0005-4019-5468}, Shigefumi Hata$^3$\orcidlink{0000-0002-6271-6982} and Hiroshi Kori$^4$\orcidlink{0000-0003-2899-7896}}

\affil{$^1$Faculty of Design, Kyushu University, Shiobaru, Fukuoka, 815-8540, Japan}

\affil{$^2$Education and Research Center for Mathematical and Data Science, Kyushu University, Motooka, Fukuoka, 819-0395, Japan}

\affil{$^3$Graduate School of Science and Engineering, Kagoshima University, Kagoshima, 890-0065, Japan}

\affil{$^4$Department of Complexity Science and Engineering, The University of Tokyo, Kashiwa, Chiba 277-8561, Japan}

\affil{$^*$Author to whom any correspondence should be addressed.}

\email{mitsui@design.kyushu-u.ac.jp}


\begin{abstract}
The dynamical order of self-sustained oscillators is often characterized by phase synchronization, extensively studied within the framework of the Kuramoto model. It has recently been reported that strong coupling leads to further organization of coupled oscillators, termed waveform proportionality (WP), through amplitude dynamics that cannot be addressed using the Kuramoto model. A previous study [Phys. Rev. Lett. 134, 167202 (2025)] showed that, in coupled oscillator systems, synchronization induces Taylor's law (TL). Particularly, it demonstrated that strong coupling gives rise to WP, which leads to TL with an exponent 2. The findings suggested that WP requires the individual oscillators constituting the coupled system to possess sufficiently fast intrinsic frequencies. Here, we show that WP and TL with an exponent 2 can be induced by a pacemaker oscillator, regardless of the magnitude of the intrinsic frequencies of the individual oscillators in a population. Specifically, even in a population composed of oscillators with slow intrinsic frequencies, WP and TL with an exponent 2 can be induced by coupling the population to a fast pacemaker. Furthermore, we demonstrate that WP and TL can also be induced in a population of non-self-oscillatory units by coupling them to a pacemaker. These results indicate that WP and TL with an exponent 2 are more universal than previously thought, extending beyond oscillator populations with fast intrinsic dynamics.
\end{abstract}

\section{Introduction}
\subsection{Collective phenomena and Kuramoto model}
The synchronization of microscopic rhythms is ubiquitously observed in the real world \cite{synchronization_book1, synchronization_book2, Kuramoto_book}. Representative examples include the periodic contraction of cardiac muscle cells \cite{Current_SAN_JTB1996, mech_cardiac_Natphys2016, Nlin_heart_review2016}, population-level firing of neurons \cite{neuro_osci_Science2004, dynamic_brain_review2008}, and aligned rotational dynamics of turbines in power grids \cite{overview_grid_power_IEEE2006, spontaneous_power_grid_NatPhys2013}.
The emergent dynamics exhibited by such populations have attracted considerble attention across diverse research fields \cite{synchronization_book1, synchronization_book2}, contributing to the understanding of fundamental biological processes and the development of engineering applications.
Importantly, synchronization also plays a critical role in ecological contexts. Synchronization among individuals, such as in the collective flashing of fireflies \cite{sync_fireflies_Nature1966, selforga_fireflies_SciAdv2021}, chorusing of frogs \cite{spatiotemporal_frog_SciRep2014}, and the spatial synchrony of plant communities and animal populations \cite{mast_seeding_Nature1998, Lynx_10year_JAE1942, sheep_sync_Nature1998, spatial_sync_review_2004}, can influence population stability, resource distribution, and ecosystem resilience. To elucidate the mechanisms underlying these synchronous behaviors, theoretical studies commonly model individual units as nonlinear oscillators and investigate the population dynamics of their coupled systems.

Synchronization is a collective behavior in which microscopic rhythms become aligned; therefore, the phase of the oscillations is essential for describing the state of each oscillator. The Kuramoto model describes the state of an oscillator solely in terms of its phase, which increases with a constant natural frequency \cite{Kuramoto_book}.
The elements are coupled through sinusoidal coupling functions and interact to adjust their phases. This established model has successfully accounted for the synchronization phenomena universally observed across different systems.
Remarkably, the Kuramoto model is derived by applying the phase reduction method to general limit-cycle oscillators with diffusive coupling \cite{Kuramoto_book, Nakao_reduction_review}. This implies that under certain conditions, the collective behavior of coupled oscillators can be approximately described by the Kuramoto model, regardless of the detailed design of the individual elements. This theoretically supports that synchronization is a universal phenomenon occurring across a wide range of objects and scales.

A key assumption in phase reduction is that the interactions between oscillators are sufficiently weak and that the modulation of the oscillation amplitude of individual units owing to these interactions is negligible. Therefore, the collective dynamics of strongly coupled oscillators may deviate from this framework and 
cannot be fully captured using phase dynamics alone. 
Examples include amplitude death, oscillation death, and oscillation quenching \cite{OQ_review_2012, OQ_review_2013, OQ_review_2021}. These phenomena represent the cessation of oscillations in individual units caused by coupling, and their analysis requires consideration of the amplitude degrees of freedom. Another example is chaotic phase synchronization \cite{Rosenblum_etal_1996_PRL, Pikovsky_etal_1996_EPL}. In systems of coupled chaotic oscillators, phase synchronization can occur, in which the phases become aligned while the amplitudes remain uncorrelated. Although phase dynamics plays a central role in analyzing chaotic phase synchronization, the amplitude degrees of freedom are also essential. Moreover, generalized synchronization, in which two time series $\bm{x}(t)$ and $\bm{y}(t)$ satisfy a functional relation such as $\bm{y}(t) = \bm{F}[\bm{x}(t)]$ \cite{Rulkov_etal_1995_PRE, Kocarev_Parlitz_1996_PRL}, and projective synchronization, in which the time series differ only by a constant factor \cite{Mainieri_Rehacek_1999_PRL}, are collective phenomena of coupled oscillators that cannot be captured by phase-only descriptions.

Recently, it has been shown that in several coupled oscillator models, including coupled Rössler systems, exhibit a synchronous state in which the time series become proportional to one another, referred to as waveform proportionality (WP) \cite{Mitsui_Kori_2025_PRL, Mitsui_Kori_arXiv2025}. As a consequence of WP, Taylor's law (TL) with an exponent 2 arises naturally \cite{Mitsui_Kori_2025_PRL, Mitsui_Kori_arXiv2025}. TL refers to the power-law relationship between the mean and variance \cite{Bliss_1941, Taylor_1961_Nature}; details are presented in the following section.

\subsection{Taylor's law}
TL is the scaling relationship between the mean and variance of a measurement \cite{Bliss_1941, Taylor_1961_Nature}. TL has been documented across a broad range of research fields and has been studied intensively in ecology, where it was originally discovered \cite{Eisler_etal_2008_AP, Taylor_2019_book}. In physics, TL is also known as fluctuation scaling. Depending on the value of the exponent, TL is sometimes referred to as giant (number/density) fluctuations or hyperuniformity \cite{GNF_localHU_PRE2025}. When the exponent is greater than one, TL can be referred to as giant fluctuations, giant number fluctuations, or giant density fluctuations \cite{GNF_Ginelli_2016, GNF_Chate_2019, Nishiguchi_2023_JPSJ}, whereas when the exponent is less than one, TL can be referred to as hyperuniformity \cite{HU_review_2018}.

TL is typically classified into two types depending on how the mean and variance are computed. When we use temporal means and variances for each time series, the scaling relationship is referred to as temporal TL. In contrast, when we use ensemble means and variances at each time point, it is referred to as spatial TL. Mathematically, TL can be expressed as follows:
\begin{align}
\log(\mathrm{variance}) = \log \alpha_{\mathrm{t,s}} + \beta_{\mathrm{t,s}} \times \log(\mathrm{mean}),
\end{align}
where $\log \alpha_{\mathrm{t}}$ ($\log \alpha_{\mathrm{s}}$) denotes the intercept of temporal (spatial) TL, and $\beta_{\mathrm{t}}$ ($\beta_{\mathrm{s}}$) denotes the exponent of temporal (spatial) TL.

Although theoretical studies have shown that the TL exponent can take arbitrary values \cite{Cohen_etal_2013_PRSB, Cohen_2014_TE, Cohen_2014_TPB}, emprical data in ecology often yield exponents close to 2 \cite{Taylor_Woiwod_1982_JAE, Kerkhoff_Ballantyne_2003_EL, Zhao_etal_2019_JAE}. This discrepancy between theoretical predictions and empirical observations has motivated the development of theories that yield TL with an exponent 2 \cite{Mitsui_Kori_2025_PRL, Mitsui_Kori_arXiv2025, Giometto_etal_2015_PNAS}. Previous studies have shown that the TL exponent approaches 2 as the correlation between the time series increases \cite{Kerkhoff_Ballantyne_2003_EL, Ballantyne_Kerkhoff_2005_JTB, Hanski_1987}. Moreover, Reuman \textit{et al}. demonstrated that spatial TL with an exponent 2 arises when correlations between time series become sufficiently strong that the time series are proportional to each other \cite{Reuman_etal_2017_PNAS}.

Motivated by these studies, Mitsui and Kori hypothesized that synchronization may underlie the emergence of TL with an exponent 2 \cite{Mitsui_Kori_2025_PRL, Mitsui_Kori_arXiv2025}. Using coupled oscillator models, they tested this hypothesis and found that strong coupling leads to WP, a synchronous state in which the time series become proportional to each other, and showed that both temporal and spatial TLs with an exponent 2 emerge as a consequence of this state \cite{Mitsui_Kori_2025_PRL, Mitsui_Kori_arXiv2025}.

 According to the theory developed in \cite{Mitsui_Kori_2025_PRL}, WP requires that the individual oscillators constituting the population possess sufficiently fast intrinsic frequencies. Here, we relax this condition and show that WP and TL with an exponent 2 can arise even when not all the oscillators possess fast intrinsic frequencies. We consider a system composed of uncoupled oscillators driven by a common input from a pacemaker oscillator. This setting enables a straightforward investigation of the transition from synchronization to the emergence of TL because the synchronization in this system is achieved simply through phase locking of the individual elements to the pacemaker. We develop a theory to account for the emergence of TL under synchronized conditions. Based on this theory, we demonstrate that WP and TL with an exponent 2 can arise even when not all oscillators possess fast intrinsic frequencies, provided that the pacemaker oscillator is sufficiently fast and that the oscillator population is strongly coupled to the pacemaker. Furthermore, we show that WP and TL with an exponent 2 can also be induced in a population of non-self-oscillatory units driven by a pacemaker oscillator. These results suggest that WP and TL are more general phenomena than previously thought.

The remainder of this paper is organized as follows. In Sec.~2, we provide the background for this study and define the relevant quantities. In Sec.~3, we present our results, including analytical calculations and numerical simulations. In Sec.~4, we discuss the implications of the findings.

\section{Background and definitions}
This section provides detailed definitions of WP and TL in coupled oscillator systems. In Ref.~\cite{Mitsui_Kori_2025_PRL}, several coupled oscillator models including the following coupled R\"ossler system\cite{Rosenblum_etal_1996_PRL, Rosenblum_etal_1997_PRL, Sakaguchi_2000_PRE, Pikovsky_etal_1996_EPL,Montbrio_Blasius_2003_Chaos, Osipov_etal_1997_PRE, locking_based_PRL2002, three_transition_PRL2003} were used:
\begin{subequations}
\label{eq:coupled_Rossler}
\begin{eqnarray}
\dot{x}_i &=& -\omega_iy_i - z_i + \dfrac{D}{N}\sum_{j=1}^{N}(x_j - x_i), \label{eq:coupled_Rossler_dot_x} \\
\dot{y}_i &=& \omega_ix_i + a y_i + \dfrac{D}{N}\sum_{j=1}^{N}(y_j - y_i), \label{eq:coupled_Rossler_dot_y} \\
\dot{z}_i &=& b +  z_i(x_i-c),\label{eq:coupled_Rossler_dot_z}
\end{eqnarray}
\end{subequations}
where $a, b,$ and $c$ are the standard parameters of the R\"ossler system \cite{Rossler1976}. Parameter $\omega_i$ encodes the approximate intrinsic frequency of each R\"ossler oscillator, and $D$ represents the coupling strength.

Since TL is typically defined for variables that take only positive values, we focus on the dynamics and TL of variable $z_i(t)$. For $z_i(t)$,  the temporal mean and variance are defined as follows:
\begin{subequations}
\label{eq:temporal_mean_var}
\begin{eqnarray}
\mathrm{E}[z_i(t)]_t &=& \langle z_i(t)\rangle_t, \\
\mathrm{Var}[z_i(t)]_t &=& \langle \left( z_i(t) - \mathrm{E}[z_i(t)]_t \right)^2 \rangle_t.
\end{eqnarray}
\end{subequations}
Here, $\langle \cdot \rangle_t$ denotes a long-time average or one-period average if the dynamics is periodic. We also define the ensemble mean and variance as
\begin{subequations}
\label{eq:ensemble_mean_var}
\begin{eqnarray}
\mathrm{E}[z_i(t)]_i &=& \langle z_i(t)\rangle_i, \\
\mathrm{Var}[z_i(t)]_i &=& \langle \left( z_i(t) - \mathrm{E}[z_i(t)]_i \right)^2 \rangle_i.
\end{eqnarray}
\end{subequations}
Here, $\langle \cdot \rangle_i$ denotes the ensemble average of all oscillators. 

\begin{figure}[tb]
 \centering
        \includegraphics[width=120mm]{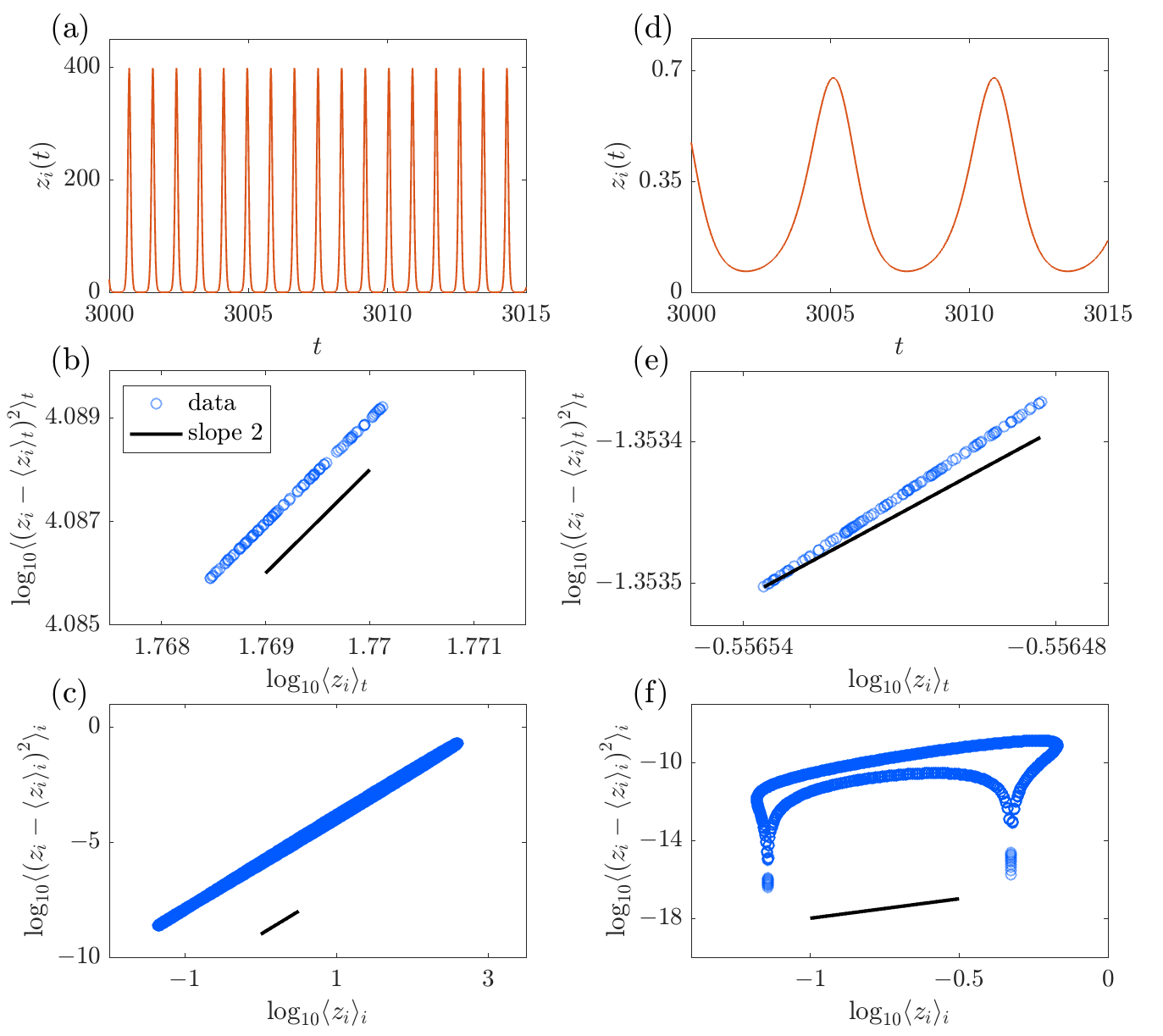}
 \caption{Examples of synchronized time series and TL in the globally coupled R\"ossler system. $N=100, D = 1000, a = 0.1$, $b=0.1$, and $c=0.7$. In panels (a)--(c), $\omega_i$ is randomly drawn from a uniform distribution between $5.4$ and $5.6$. In panels (d)--(f), $\omega_i$ is randomly drawn from a uniform distribution between $0.9$ and $1.1$. In panels (a) and (d), although the 100 time series appear as a single trace, they are not completely synchronized actually. (a), (d) Time series of $z_i(t)$. (b), (e) Scatter plot for temporal mean and variance of $z_i(t)$. (c), (f) Scatter plot for ensemble mean and variance of $z_i(t)$.}
\label{fig:coupled_Rossler_timeseries_TL}
\end{figure}

We now introduce WP, which can generally be expressed as
\begin{eqnarray}
z_i(t) = C_i z_0(t - t_i), \label{eq:WP_with_lag}
\end{eqnarray}
where $C_i$ is a constant for each oscillator and $t_i$ represents the time lag from the reference dynamics $z_0(t)$, which is expected to decrease as the coupling strength $D$ increases. When the coupling is sufficiently strong, we can neglect $t_i$ to obtain
\begin{eqnarray}
z_i(t) = C_i z_0(t). \label{eq:WP}
\end{eqnarray}
Note that when each oscillator has sufficiently fast dynamics, the relation~\eqref{eq:WP} can be derived from the system~\eqref{eq:coupled_Rossler} under strong coupling conditions \cite{Mitsui_Kori_2025_PRL}. Using the relation~\eqref{eq:WP}, the temporal mean, temporal variance, ensemble mean, and ensemble variance of $z_i(t)$ are given by
\begin{eqnarray}
        {\rm E}[z_i(t)]_t &=& C_i\, {\rm E}[z_0(t)]_t, \label{eq:temporal_mean_z_i}\\
        {\rm Var}[z_i(t)]_t &=& C_i^{\, 2}\, {\rm Var}[z_0(t)]_t, \label{eq:temporal_var_z_i} \\
        {\rm E}[z_i(t)]_i &=& {\rm E}[C_i]_i\, z_0(t), \label{eq:ensemble_mean_z_i}\\
        {\rm Var}[z_i(t)]_i &=& {\rm Var}[C_i]_i\, \left[z_0(t)\right]^2. \label{eq:ensemble_var_z_i}
\end{eqnarray}
From these expressions, we obtain the following relationships between the mean and variance of $z_i(t)$:
\begin{align}
\mathrm{Var}[z_i(t)]_t &= \dfrac{\mathrm{Var}[z_0(t)]_t}{\mathrm{E}[z_0(t)]_t^{\, 2}}\, \mathrm{E}[z_i(t)]_t^{\, 2}, \label{eq:temporalTL} \\
\mathrm{Var}[z_i(t)]_i &= \dfrac{\mathrm{Var}[C_i]_i}{\mathrm{E}[C_i]_i^{\, 2}}\, \mathrm{E}[z_i(t)]_i^{\, 2}. \label{eq:spatialTL}
\end{align}
The first relationship represents temporal TL with an exponent 2, whereas the second represents spatial TL with an exponent 2. Figure~\ref{fig:coupled_Rossler_timeseries_TL} shows the typical dynamics of $z_i(t)$ and the relationship between its mean and variance observed in the system \eqref{eq:coupled_Rossler} for sufficiently large $D$.
As shown in the figure, all oscillators synchronize well regardless of whether thier intrinsic dynamics are fast [Fig.~\ref{fig:coupled_Rossler_timeseries_TL}(a)] or slow [Fig.~\ref{fig:coupled_Rossler_timeseries_TL}(d)].
However, as manifested in a previous study~\cite{Mitsui_Kori_2025_PRL}, TL with an exponent 2 emerges only in the former case [Figs.~\ref{fig:coupled_Rossler_timeseries_TL}(b) and (c)] and not in the latter case [Figs.~\ref{fig:coupled_Rossler_timeseries_TL}(e) and (f)]. Thus, although TL may emerge through synchronization, synchronization alone is not sufficient to account for TL. 

The relationship among synchronization, WP, and TL are discussed in detail later in this paper. For temporal TL, a linear fitting to $N$ data points of $(\log {\rm E}[z_i(t)]_t, \log {\rm Var}[z_i(t)]_t)$ yields the slope $\beta_{\rm t}$ and intercept $\log \alpha_{\rm t}$. For spatial TL, a linear fitting to $M$ data points $(\log {\rm E}[z_i(t)]_i, \log {\rm Var}[z_i(t)]_i)$, where $M$ is the number of sample times, yields the slope $\beta_{\rm s}$ and intercept $\log \alpha_{\rm s}$. The coefficient of determination for linear fitting, quantifying the degree of TL emergence, is defined as
\begin{eqnarray}
    R^2 = \left(\dfrac{\mbox{E}\left[(X-\mbox{E}[X]) (Y-\mbox{E}[Y])\right]}{\sqrt{\mbox{E}\left[(X-\mbox{E}[X])^2\right] \mbox{E}\left[(Y-\mbox{E}[Y])^2\right]}}\right)^2,
\end{eqnarray}
where $X = \log\mbox{E}[z_i(t)]_t$ and $Y = \log\mbox{Var}[z_i(t)]_t$ in the case of temporal TL, whereas $X = \log\mbox{E}[z_i(t)]_i$ and $Y = \log\mbox{Var}[z_i(t)]_i$ in the case of spatial TL. To quantify the degree of synchronization within the oscillator population, we introduce the Kuramoto order parameter $\eta(t)$:
\begin{eqnarray}
  \eta(t) = \dfrac{1}{N}\left| \sum_{j=1}^{N}e^{\sqrt{-1}\theta_j(t)} \right|,
\end{eqnarray}
where $\theta_i(t)$, the phase of each oscillator, is defined as
\begin{eqnarray}
   \theta_i(t) = \arctan\left(\dfrac{y_i(t)-\langle y_i(t) \rangle_t}{x_i(t)-\langle x_i(t) \rangle_t}\right).
\end{eqnarray}
$\eta(t)$ takes values in the range of 0 to 1. When the phases of the oscillators are uniformly distributed, $\eta(t) = 0$, and when they have the same phase, $\eta(t) = 1$. In this study, we use the time-averaged Kuramoto order parameter, $\langle \eta(t) \rangle_t$, to evaluate the degree of synchronization. Note that, in previous studies, the following order parameter $\chi$ was used \cite{Mitsui_Kori_2025_PRL, Mitsui_Kori_arXiv2025}:
\begin{eqnarray}
\chi = \dfrac{\mbox{CV}[Z(t)]}{\underset{i}{\max}\{\mbox{CV}[z_i(t)]\}},\label{eq:chi}
\end{eqnarray}
where $Z(t) = \langle z_i(t) \rangle_i$ and CV represents the coefficient of variation. In other words, $\chi$ is the CV of the mean field of $z_i(t)$ normalized by the maximum CV of the individual $z_i(t)$. By defining $\chi$ in this manner, $\chi$ takes values in the range of 0 to 1. $\chi$ exhibits similar behavior to that of $\langle \eta(t) \rangle_t$. Both $\langle \eta(t) \rangle_t$ and $\chi$ quantify the degree of the phase synchronization. While the results obtained in this study are insensitive to which order parameter is used, we use $\langle \eta(t) \rangle_t$ here. Although $\langle \eta(t) \rangle_t$ is computed from $\theta_i(t)$, which is defined using $x_i(t)$ and $y_i(t)$, it should be noted that $\langle \eta(t) \rangle_t$ nevertheless properly reflects the degree of phase synchronization of $z_i(t)$ as well (see Fig.~\ref{fig:eta_chi} in Appendix).

\section{Results}
In this section, we present the results of a population of uncoupled oscillators driven by a pacemaker. For the numerical simulations, the population dynamics is calculated up to $t = 3500$, and TL is computed using the time series from $t = 3000$ to $t = 3500$. The error bars shown in each figure represent the standard deviation obtained from ten simulations with different initial conditions and realizations of the parameter $\omega_i$. For each oscillator, the initial conditions are drawn from a uniform distribution between $0$ and $1$. When the intrinsic frequencies of the pacemaker oscillator and the oscillator population differ substantially, the time series of the oscillator population may diverge at certain coupling strengths. The results of these cases are excluded.

\subsection{Pacemaker-induced Taylor's law with an exponent 2}

In this study, we show that introducing a pacemaker oscillator can induce TL with an exponent 2 through synchronization, even in a population of oscillators with slow intrinsic frequencies. To this end, we consider the following system, in which each oscillator is coupled to a pacemaker oscillator:
\begin{subequations}
\label{eq:pacemaker_Rossler}
\begin{eqnarray}
\dot{x}_{\mathrm p} &=& -\omega_{\mathrm p}y_{\mathrm p} - z_{\mathrm p}, \label{eq:pacemaker_Rossler_dot_xp} \\
\dot{y}_{\rm p} &=& \omega_{\mathrm p}x_{\mathrm p} + a y_{\mathrm p}, \label{eq:pacemaker_Rossler_dot_yp} \\
\dot{z}_{\rm p} &=& b +  z_{\mathrm p}(x_{\mathrm p}-c), \label{eq:pacemaker_Rossler_dot_zp}\\
\dot{x}_i &=& -\omega_iy_i - z_i + D(x_{\mathrm p} - x_i), \label{eq:pacemaker_Rossler_dot_x} \\
\dot{y}_i &=& \omega_ix_i + a y_i + D(y_{\mathrm p} - y_i), \label{eq:pacemaker_Rossler_dot_y} \\
\dot{z}_i &=& b +  z_i(x_i-c),\label{eq:pacemaker_Rossler_dot_z}
\end{eqnarray}
\end{subequations}
where $x_{\mathrm p}, y_{\mathrm p},$ and $z_{\mathrm p}$ are the variables of the pacemaker oscillator, and $x_i, y_i,$ and $z_i \ (i=1, \cdots, N)$ are the variables that describe the dynamics of oscillator $i$ connected to the pacemaker oscillator. $\omega_{\mathrm p}$ and $\omega_i$ represent the approximate intrinsic frequencies of the pacemaker oscillator and each oscillator $i$ in the population, respectively.

\begin{figure}[tb]
 \centering
        \includegraphics[width=130mm]{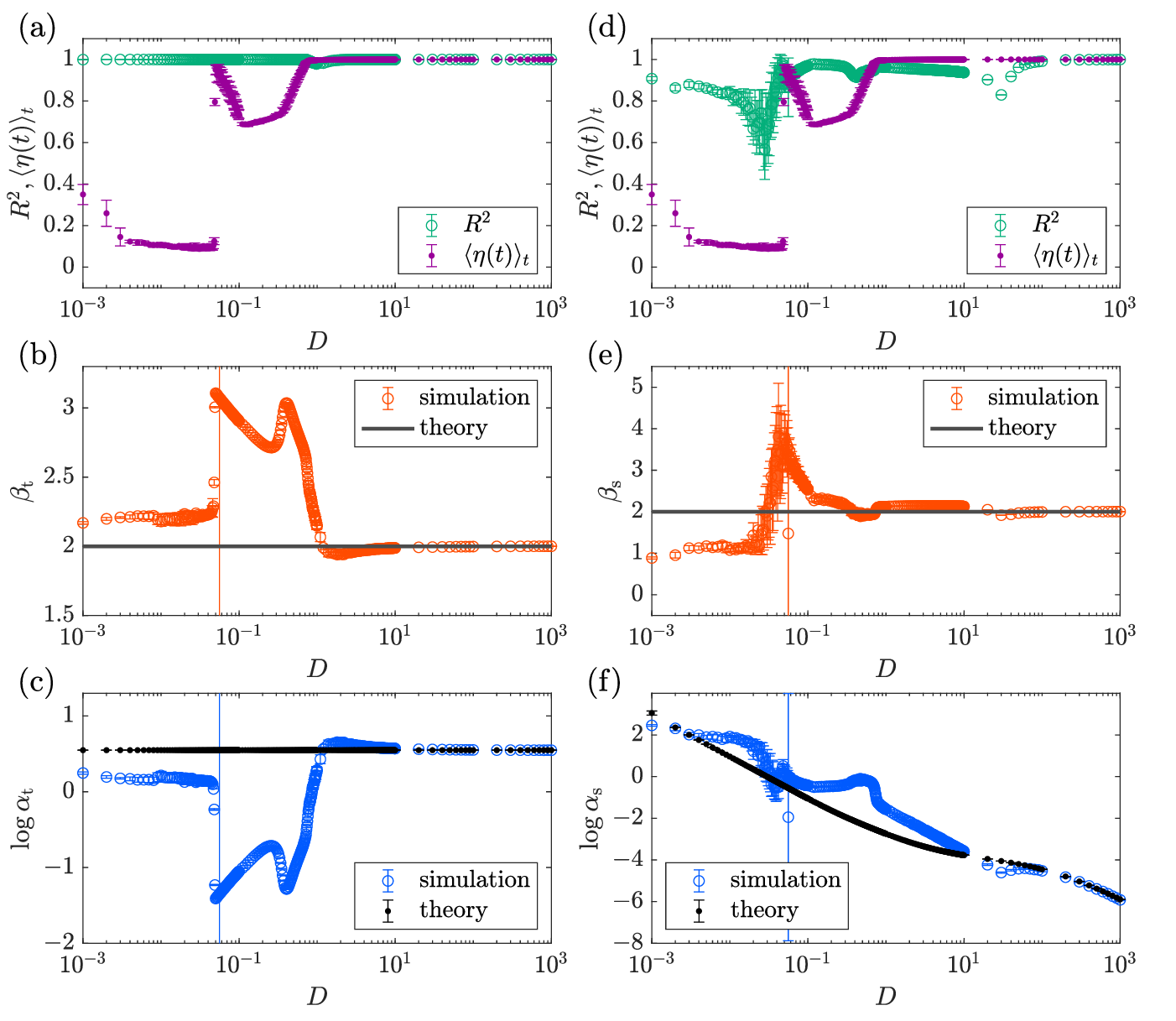}
 \caption{Dependence of TL parameters and synchronization degree on $D$ in the pacemaker-driven R\"ossler system. $N=100, \omega_{\mathrm p}=5.5, a = 0.1$, $b=0.1$, and $c=0.7$. $\omega_i$ is randomly drawn from a uniform distribution between $5.4$ and $5.6$. (a) The coefficient of determination, $R^2$, for temporal TL and the time-averaged Kuramoto order parameter $\langle \eta(t) \rangle_t$. (b) Exponents of temporal TL. (c) Intercepts of temporal TL. (d) The coefficient of determination, $R^2$, for spatial TL and the time-averaged Kuramoto order parameter $\langle \eta(t) \rangle_t$. (e) Exponents of spatial TL. (f) Intercepts of spatial TL. The theoretical predictions of $\log\alpha_{\mathrm t, s}$ are obtained by setting $\tau = 0.168$ [see Eqs.~\eqref{eq:ansatz}--\eqref{eq:intercept} for details].}
\label{fig:pacemaker_Rossler_TL_vs_D_with_fast_population}
\end{figure}

First, numerical investigations are conducted using Eq.~\eqref{eq:pacemaker_Rossler} to observe the emergence of TL through synchronization (Fig.~\ref{fig:pacemaker_Rossler_TL_vs_D_with_fast_population}).
In this system, TL with an exponent 2 emerges in the strong-coupling regime. Although the time-averaged Kuramoto order parameter $\langle \eta(t) \rangle_t$ is close to 1 at $D \approx 0.05$, neither the temporal nor spatial TL exponent is equal to 2, suggesting that synchronization has not yet involved WP at this coupling strength. As the coupling strength increases, the exponent $\beta_{\mathrm t}$ of temporal TL fluctuates and eventually approaches $\beta_{\mathrm t} = 2$ around $D \approx 2$; the intercept $\log \alpha_{\mathrm t}$ also approaches the theoretical prediction [see the detailed derivation in Eqs.~\eqref{eq:ansatz}--\eqref{eq:intercept}]. In contrast, for spatial TL, the exponent remains different from 2, and the theoretical prediction for $\log \alpha_{\mathrm s}$ does not agree with the numerical results, indicating that WP with a phase lag, as expressed by Eq.~\eqref{eq:WP_with_lag}, still remains. With a further increase in the coupling strength, around $D \approx 100$, the theoretical predictions and numerical results of both the exponent and intercept of spatial TL come into good agreement, suggesting that WP described by Eq.~\eqref{eq:WP} is established. This result is consistent with the previous results obtained using a pacemaker-driven food chain model \cite{Mitsui_Kori_2025_PRL}.

\begin{figure}[tb]
 \centering
        \includegraphics[width=120mm]{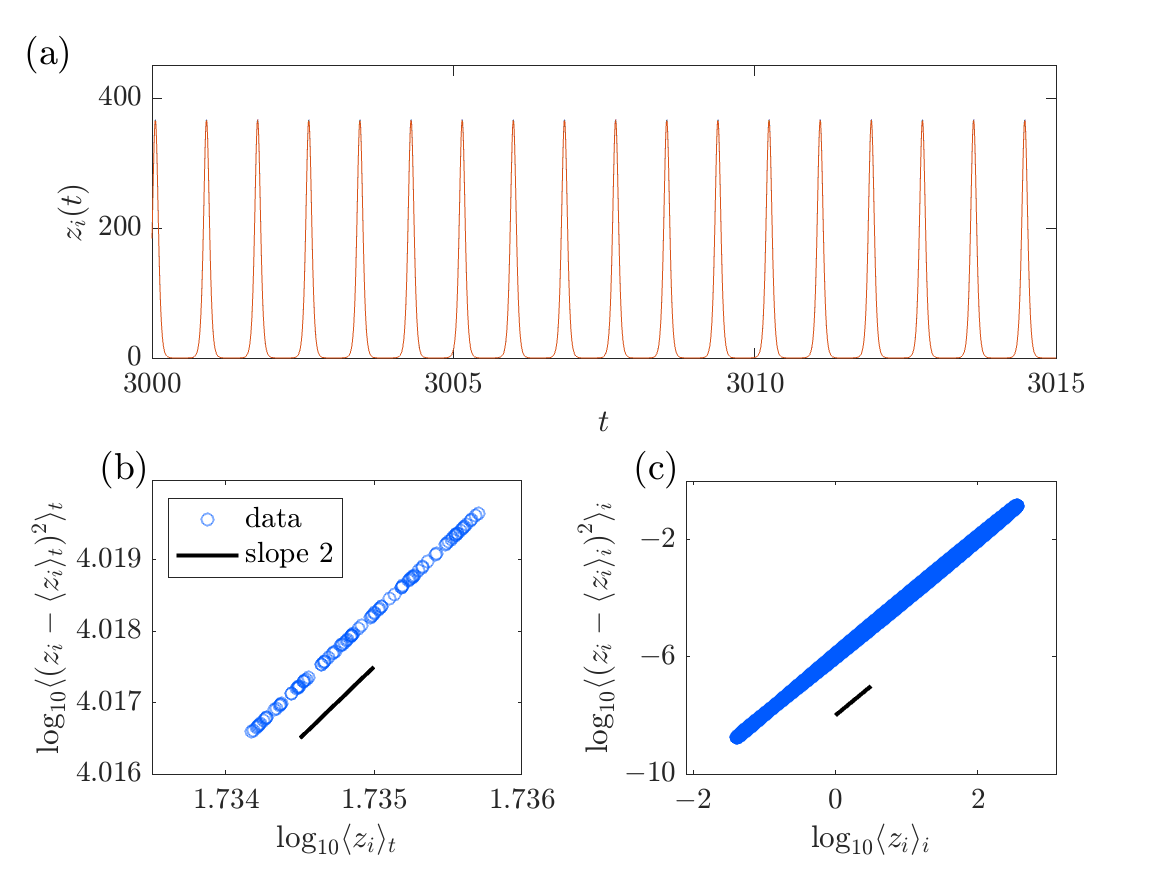}
 \caption{Examples of synchronized time series and TL with an exponent 2 in the pacemaker-driven R\"ossler system. $N=100, D = 1000, a = 0.1$, $b=0.1$, $c=0.7$, and $\omega_{\mathrm p} = 5.5$. $\omega_i$ is randomly drawn from a uniform distribution between $0.9$ and $1.1$. (a) Time series of $z_i(t)$. Although the 100 time series appear as a single trace, they are not completely synchronized actually. (b) Scatter plot of the temporal mean and variance of $z_i(t)$. (c) Scatter plot of the ensemble mean and variance of $z_i(t)$.}
\label{fig:example_pacemaker_Rossler_TL}
\end{figure}

In Ref.~\cite{Mitsui_Kori_2025_PRL}, through analytical calculations and numerical simulations, it was suggested that when individual oscillators possess fast intrinsic frequencies, both temporal and spatial TLs can emerge. Here, we extend these results and show that even when the intrinsic dynamics of the oscillator population are slow, TL with an exponent 2 emerges as long as the population is coupled to a fast pacemaker oscillator. 
Figure~\ref{fig:example_pacemaker_Rossler_TL} shows TL with an exponent 2 observed in the slow oscillator population driven by a fast pacemaker. In this example, we choose a large value of $\omega_{\mathrm p}$ for the pacemaker oscillator that produces sufficiently fast oscillations, whereas the remaining oscillators are given small values of $\omega_i$ such that their intrinsic dynamics are slow. The distribution of $\omega_i$ is chosen to be the same as that used in Figs.~\ref{fig:coupled_Rossler_timeseries_TL}(d)--(f); that is, a parameter regime where TL with an exponent 2 does not emerge on its own. We find that TL with an exponent 2 emerges even when the intrinsic dynamics of the non-pacemaker oscillators are slow, provided that they are strongly coupled to the pacemaker oscillator with sufficiently fast dynamics.

 The mechanism underlying this phenomenon can be understood analytically in the same manner as in Ref.~\cite{Mitsui_Kori_2025_PRL}, using perturbative calculations and an averaging method. Following Ref.~\cite{Mitsui_Kori_2025_PRL}, we consider the following ansatz:
\begin{subequations}
  \label{eq:ansatz}
\begin{align}
 x_i(t) &= x_{\mathrm p}(t- \eps_i \tau) + \eps_i p(t- \eps_i \tau) + O(\eps_i^{\, 2}), \label{x_ansatz}\\
 y_i(t) &= y_{\mathrm p}(t- \eps_i \tau) + \eps_i q(t- \eps_i \tau) + O(\eps_i^{\, 2}), \label{y_ansatz}\\
 z_i(t) &= z_{\mathrm p}(t- \eps_i \tau) + \eps_i r(t- \eps_i \tau) + O(\eps_i^{\, 2}), \label{z_ansatz}
\end{align}
\end{subequations}
where $p(t)$, $q(t)$, and $r(t)$ are the functions to be determined and $\tau$ is a constant. $\eps_i$ is a small parameter  defined as
\begin{subequations}
\label{eq:def_eps}
\begin{eqnarray}
    \mu_i &=& \omega_i - \omega_{\mathrm p},
    \label{eq:mu}\\
    \eps_i &=& \dfrac{\mu_i}{D}.
    \label{eq:epsilon}
\end{eqnarray}
\end{subequations}

It should be noted that ansatz~\eqref{eq:ansatz} assumes that all oscillators are frequency-locked by the pacemaker; no slips are observed. By substituting the ansatz~\eqref{eq:ansatz} into Eq.~\eqref{eq:pacemaker_Rossler} and extracting the $O(\eps_i)$ terms, we obtain the evolution equations to be satisfied by $p(t), q(t),$ and $r(t)$:
\begin{subequations}
\label{eq:Rossler_pqr}    
\begin{eqnarray}
    \dot{p} &=& - Dp -\omega_{\mathrm p}q - r - Dy_{\mathrm p}(1+\tau \omega_{\mathrm p})-D\tau z_{\mathrm p}, \label{eq:Rossler_dot_p} \\
    \dot{q} &=& (a - D)q + \omega_{\mathrm p}p + Dx_{\mathrm p}(1+\tau \omega_{\mathrm p}) + aD\tau y_{\mathrm p}, \label{eq:Rossler_dot_q} \\
    \dot{r} &=& (x_{\mathrm p} - c)r + pz_{\mathrm p}. \label{eq:Rossler_dot_r} 
\end{eqnarray}
\end{subequations}
We then apply the averaging method to obtain approximate expressions for $z_{\mathrm p}(t)$ and $r(t)$. For convenience, we set the following quantities:
\begin{eqnarray}
    f(t) &=& x_{\mathrm p}(t) - c,\\
    \bar f &=& \langle x_{\mathrm p}(t) - c\rangle_t,\\
   \delta f(t) &=& f(t) - \bar f,\\
   \delta F(t) &=& \int_0^{t}\delta f(t')dt'.
\end{eqnarray}
First, for $z_{\mathrm p}(t)$, assuming that $x_{\mathrm p}(t)$ is provided, integrating Eq.~\eqref{eq:pacemaker_Rossler_dot_zp} yields the following expression:
\begin{eqnarray}
    z_{\mathrm p}(t) = \left[ \kappa_1 + b\int_{0}^{t}e^{-\bar ft' - \delta F(t')}dt'\right]e^{\bar ft + \delta F(t)},
    \label{eq:zp}
\end{eqnarray}
where $\kappa_1$ is an integration constant. Note that $e^{-\delta F(t)}$ is a periodic function; thus we can expand $e^{-\delta F(t)}$ in the Fourier series as follows:
\begin{eqnarray}
    e^{-\delta F(t)} = A + \sum_{n=1}^{\infty}\left[a_n\cos(\omega n t) + b_n\sin(\omega n t)\right], \label{eq:Fourier}
\end{eqnarray}
where $\omega$ is the frequency of the dynamics, and $A$, $a_n$, and $b_n$ are the Fourier coefficients. In particular, $A$ is computed as
\begin{eqnarray}
 A = \langle e^{-\delta F(t)} \rangle_t. \label{eq:A_def}    
\end{eqnarray}
 Note that since $\omega_{\mathrm p}$ is the approximate frequency of the pacemaker oscillator, we use $\omega$ to denote the actual frequency of the observed dynamics. We substitute Eq.~(\ref{eq:Fourier}) into the integral term of Eq.~(\ref{eq:zp}) to obtain
\begin{align}
 \int_0^t e^{-\bar f t' - \delta F(t')} dt'
 =& e^{-\bar f t}
\left[
-\frac{A}{\bar f}+ \sum_{n=1}^{\infty} \dfrac{(a_n\omega n - b_n \bar f) \sin(\omega nt) - (a_n \bar f + b_n\omega n) \cos(\omega nt)}{\bar{f}^2+(\omega n)^2}\right] \nonumber \\ 
&+ \sum_{n=1}^{\infty}\dfrac{a_n\bar{f} + b_n \omega n}{\bar{f}^2+(\omega n)^2} + \dfrac{A}{\bar{f}}.
\end{align}
Substituting this expression into Eq.~(\ref{eq:zp}), we obtain
\begin{gather}
z_{\mathrm p}(t) =\left\{ \Tilde{\kappa}_1 + be^{-\bar f t}
\left[
-\frac{A}{\bar f}+ \sum_{n=1}^{\infty} \dfrac{(a_n\omega n - b_n \bar f) \sin(\omega nt) - (a_n \bar f + b_n\omega n) \cos(\omega nt)}{\bar{f}^2+(\omega n)^2}\right]\right\}e^{\bar f t + \delta F(t)},
\end{gather}
where
\begin{align}
\Tilde{\kappa}_1 = \kappa_1 + b\sum_{n=1}^{\infty}\dfrac{a_n\bar{f} + b_n\omega n}{\bar{f}^2+(\omega n)^2} + b\dfrac{A}{\bar{f}},
\end{align}
denotes a time-independent constant. Now, we set $\Tilde{\kappa}_1 = 0$ because we focus on periodic dynamics. We also assume that the oscillation is sufficiently fast so that the condition $|\bar f| \ll \omega$ is satisfied, which gives
\begin{eqnarray}
    z_{\mathrm p}(t) = \dfrac{b}{\bar f}\left[ -A + O\left(\dfrac{\bar f}{\omega}\right) \right]e^{\delta F(t)}.
\end{eqnarray}
A similar calculation can be performed for $r(t)$. Assuming that $x_{\mathrm p}(t)$, $z_{\mathrm p}(t)$, and $p(t)$ are provided, we can perform the integration of Eq.~\eqref{eq:Rossler_dot_r}. We then obtain the following expression:
\begin{eqnarray}
    r(t) = \left[ \kappa_2 + \int_{0}^{t}p(t')z_{\mathrm p}(t')e^{-\bar ft' - \delta F(t')}dt'\right]e^{\bar ft + \delta F(t)},
\end{eqnarray}
where $\kappa_2$ is an integration constant. When $|\bar f| \ll \omega$ is satisfied, we obtain
\begin{eqnarray}
    r(t) = \dfrac{1}{\bar f}\left[ -B + O\left(\dfrac{\bar f}{\omega}\right) \right]e^{\delta F(t)},
\end{eqnarray}
where $B$ is computed as
\begin{eqnarray}
    B = \langle p(t)z_{\mathrm p}(t) e^{-\delta F(t)} \rangle_t. \label{eq:B_def}    
\end{eqnarray}
 In summary, when the conditions
\begin{subequations}
\label{eq:WP_condition}
 \begin{eqnarray}
     \omega&\gg& |\bar f|,\\
    |A| &\gg& \left|\dfrac{\bar{f}}{\omega}\right|,\\ 
    |B| &\gg& \left|\dfrac{\bar{f}}{\omega}\right|,
 \end{eqnarray}
\end{subequations}
  are satisfied, we obtain the following expression, which means WP:
\begin{eqnarray}
    z_i(t) &\simeq& z_{\mathrm p}(t) + \eps_ir(t),\\
           &\simeq& -\dfrac{Ab}{\bar f}  e^{\delta F(t)} -\dfrac{\eps_iB}{\bar f}e^{\delta F(t)},\\
           &=& -\dfrac{Ab + \eps_iB}{\bar f}e^{\delta F(t)}. \label{eq:app_z_i}
\end{eqnarray}
From Eq.~\eqref{eq:app_z_i}, we straightforwardly obtain
\begin{subequations}
 \label{eq:intercept}
    \begin{eqnarray}
    \log \alpha_{\mathrm t} &=& \log \dfrac{\mathrm{Var}[e^{\delta F(t)}]_t}{\mathrm{E}[e^{\delta F(t)}]_t^{\ 2}},\\
    \log \alpha_{\mathrm s} &=& \log \left( \dfrac{B}{Ab} \right)^2\mathrm{Var}[\eps_i]_i^{\, 2}.
\end{eqnarray}
\end{subequations}

\begin{figure}[tb]
 \centering
        \includegraphics[width=140mm]{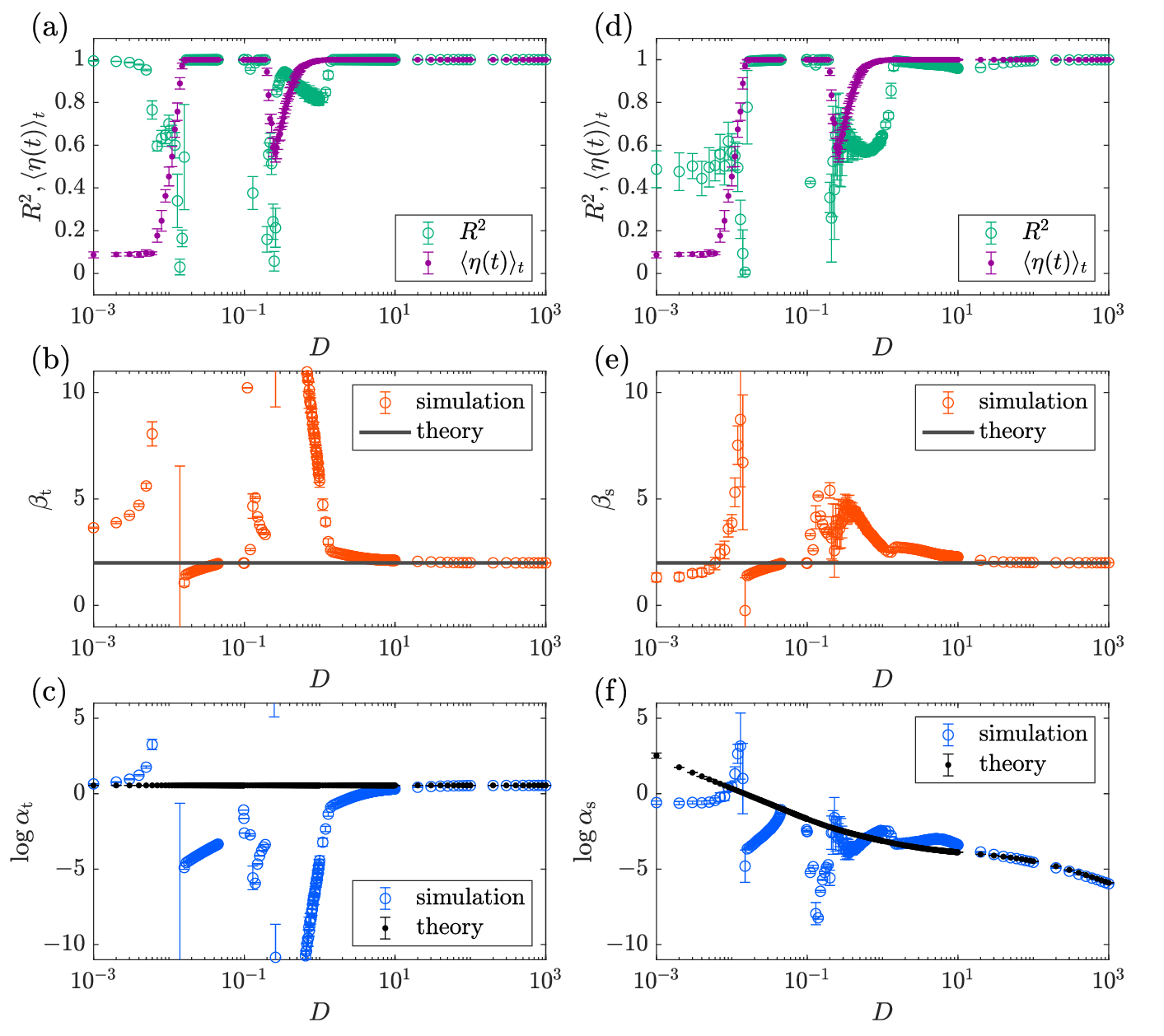}
 \caption{Dependence of TL parameters and synchronization degree on $D$ in the pacemaker-driven R\"ossler system. $N=100, \omega_{\mathrm p}=5.5, a = 0.1$, $b=0.1$, $c=0.7$, and $\tau = 0.0199$. $\omega_i$ is randomly drawn from a uniform distribution between $0.9$ and $1.1$. (a) The coefficient of determination, $R^2$, for temporal TL and the time-averaged Kuramoto order parameter $\langle \eta(t) \rangle_t$. (b) Exponents of temporal TL. (c) Intercepts of temporal TL. (d) The coefficient of determination, $R^2$, for spatial TL and the time-averaged Kuramoto order parameter $\langle \eta(t) \rangle_t$ . (e) Exponents of spatial TL. (f) Intercepts of spatial TL.}
\label{fig:pacemaker_Rossler_TL_vs_D_with_slow_population}
\end{figure}

\begin{figure}[tb]
 \centering
        \includegraphics[width=140mm]{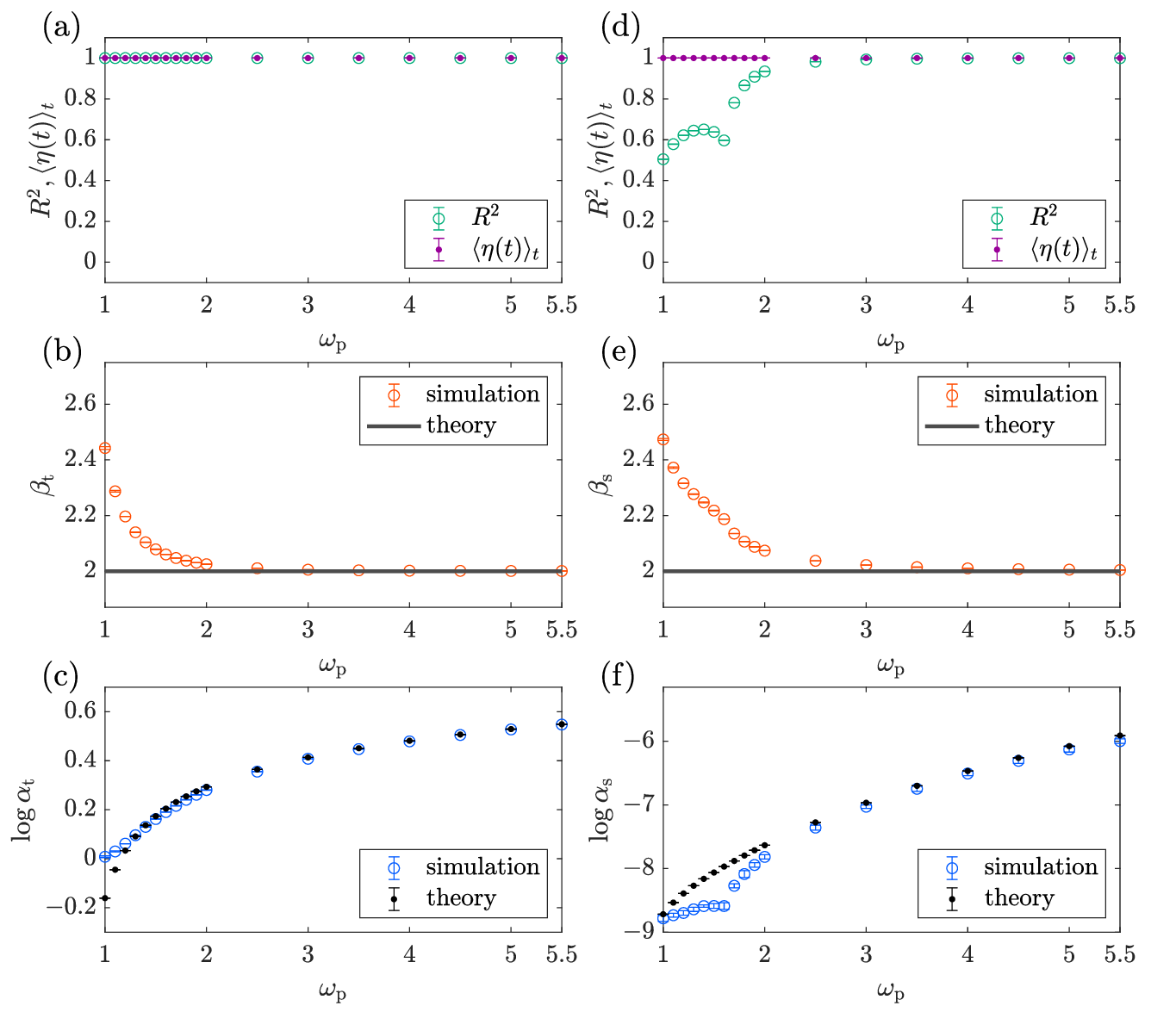}
 \caption{Dependence of TL parameters on $\omega_{\mathrm p}$ in the pacemaker-driven R\"ossler system. $N=100, D = 1000, a = 0.1$, $b=0.1$, and $c=0.7$. $\omega_i$ is randomly drawn from a uniform distribution between $0.9$ and $1.1$. Since the coupling strength is sufficiently large, we set $\tau =0$ in this numerical simulation. (a) The coefficient of determination, $R^2$, for temporal TL and the time-averaged Kuramoto order parameter $\langle \eta(t) \rangle_t$. (b) Exponents of temporal TL. (c) Intercepts of temporal TL. (d) The coefficient of determination, $R^2$, for spatial TL and the time-averaged Kuramoto order parameter $\langle \eta(t) \rangle_t$. (e) Exponents of spatial TL. (f) Intercepts of spatial TL.}
\label{fig:pacemaker_Rossler_TL_vs_omega_p}
\end{figure}

Based on the above analysis, we expect that WP will be established and therefore both temporal and spatial TLs with an exponent 2 will emerge when the oscillator population is strongly coupled to and locked with the pacemaker oscillator, even if the intrinsic dynamics of each oscillator that constitutes the population is slow. We numerically confirm this expectation. Figure~\ref{fig:pacemaker_Rossler_TL_vs_D_with_slow_population} shows the results for an oscillator population with slow intrinsic frequencies coupled to a pacemaker oscillator with a fast intrinsic frequency. The parameter $\tau$ used in the theoretical predictions of $\log \alpha_{\rm{t, s}}$ is obtained numerically by sorting the peak times of each $z_i(t)$ according to $\eps_i$ and performing linear fitting (see Appendix for details). As expected, in the strong-coupling regime, the oscillators are well synchronized, and temporal and spatial TLs with an exponent 2 emerge. Our theoretical analysis also suggests that WP and TL do not emerge when the pacemaker's frequency is low and the condition \eqref{eq:WP_condition} is not satisfied. Figure~\ref{fig:pacemaker_Rossler_TL_vs_omega_p} shows numerical simulation results confirming this prediction by varying the parameter $\omega_{\mathrm p}$, which approximately determines the frequency of the pacemaker oscillator. When $\omega_{\mathrm p}$ is small, the exponents and intercepts of TL obtained from the numerical simulations do not agree with the theoretical predictions. By contrast, when $\omega_{\mathrm p}$ is large, the numerical results are in excellent agreement with the theoretical predictions. 

\begin{figure}[tb]
 \centering
        \includegraphics[width=130mm]{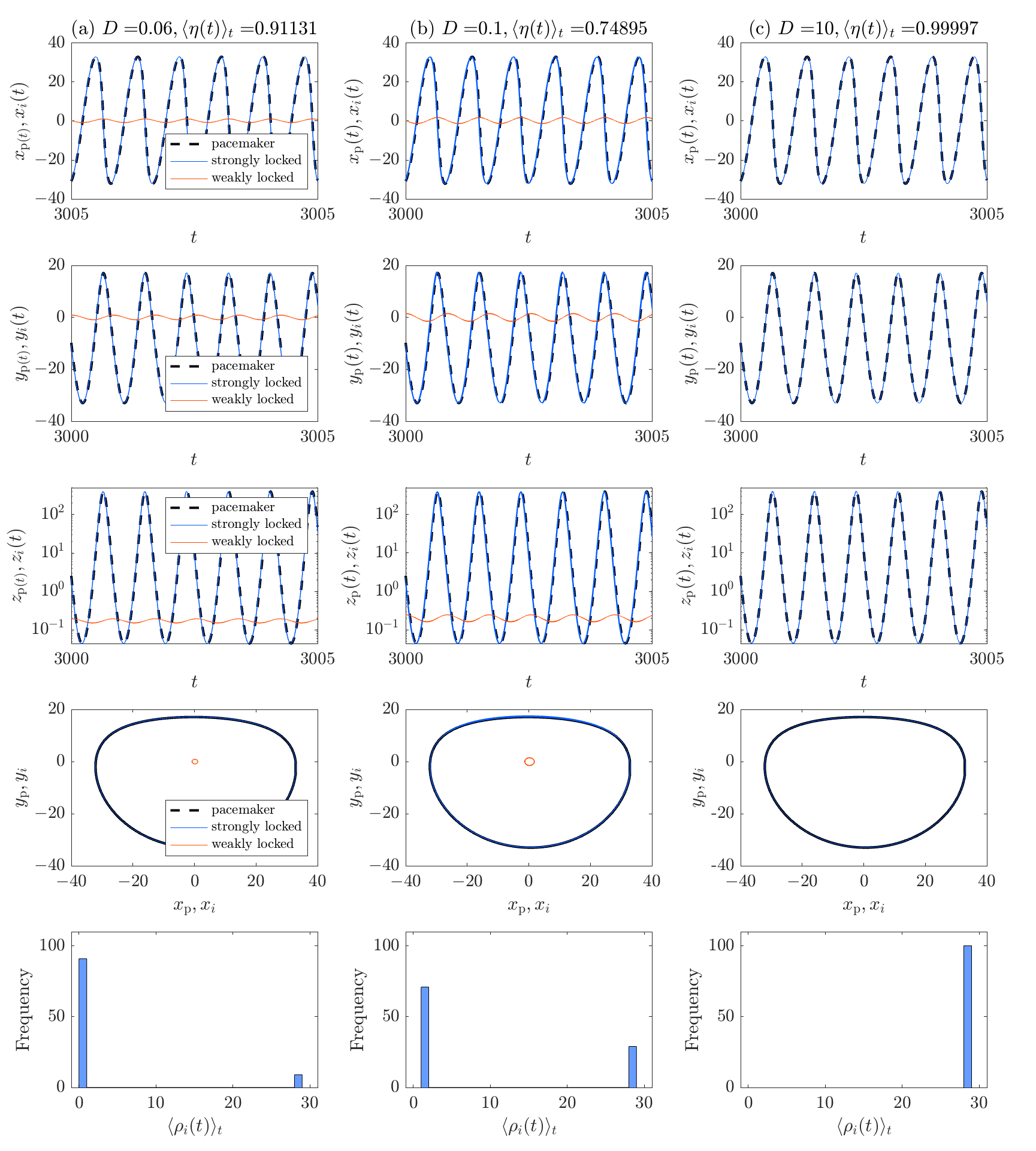}
 \caption{Time series, attractors, and amplitudes for each coupling strength. $N=100, \omega_{\mathrm p}=5.5, D = 1000$, $a = 0.1$, $b=0.1$, and $c=0.7$. $\omega_i$ is randomly drawn from a uniform distribution between $5.4$ and $5.6$. First row: time series of $x_i(t)$. Second row: time series of $y_i(t)$. Third row: time series of $z_i(t)$. Fourth row: projection of the attractors on the $x_i$-$y_i$ plane. In the first row to the forth row, dashed black lines, solid blue lines, and solid red lines represent the pacemaker oscillator, the strongly locked oscillators in a population, and the weakly locked oscillators in a population, respectively. Fifth row: distribution of the oscillator amplitudes. The oscillator amplitude is defined as $ \rho_i(t) = \sqrt{ \left(x_i(t)-\langle x_i(t) \rangle_t\right)^2 + \left(y_i(t)-\langle y_i(t) \rangle_t\right)^2}.$   (a) $D=0.06$. (b) $D=0.1$. (c) $D=10$.}
\label{fig:clustering}
\end{figure}

As the coupling strength increases, the time-averaged Kuramoto order parameter, $\langle \eta(t) \rangle_t$, exhibits intriguing behavior before the emergence of TL [Fig.~\ref{fig:pacemaker_Rossler_TL_vs_D_with_fast_population}(a) and (d)]. As is commonly observed in conventional phase oscillator models, $\langle \eta(t) \rangle_t$ gradually increases to approach $\langle \eta(t) \rangle_t \simeq 1$ ($D\simeq 0.06$) . However, when $D$ increases further, $\langle \eta(t) \rangle_t$ decreases once and has a relatively small value ($D\simeq 0.1$). Subsequently, it recovers $\langle \eta(t) \rangle_t \simeq 1$ for a sufficiently strong regime ($D\simeq 10$). This reentrant phenomenon is not generally observed in phase models; therefore, it is likely due to the amplitude effect. Indeed, we observe bistable amplitude modulation in each oscillator resulting from the locking to the pacemaker (Fig.~\ref{fig:clustering}).
The driven oscillators are locked by the pacemaker and are in frequency synchronization even when $D\simeq 0.06$, resulting in $\langle \eta(t) \rangle_t \simeq 1$ [Fig.~\ref{fig:clustering} (a)]. Note that, however, in such a weakly locked regime, each oscillator exhibits a small-amplitude oscillation that is still far from the pacemaker's state [Fig.~\ref{fig:clustering} (a)]. As the driving force of the pacemaker increases, an oscillator with an intrinsic frequency close to that of the pacemaker becomes {\it strongly} locked.
This leads to a discontinuous transition in the oscillatory amplitude to another stable regime, allowing the oscillator to closely follow the pacemaker. Oscillators with two different locking regimes therefore coexist when the coupling strength is intermediate ($D\simeq 0.1$), causing the order parameter $\langle \eta(t) \rangle_t$ to decrease once [Fig.~\ref{fig:clustering} (b)]. When the coupling becomes sufficiently strong, all oscillators become strongly locked to the pacemaker and exhibit frequency synchronization within an almost identical oscillatory orbit [Fig.~\ref{fig:clustering} (c)]. This may recover the order parameter $\langle \eta(t) \rangle_t \simeq 1$ around $D\simeq 10$ [Fig.~\ref{fig:clustering} (c)].
Note that our numerical results suggest that TL with an exponent 2 emerges when all the oscillators are in a strongly locked regime.
This is consistent with our analytical investigations, suggesting that in addition to the ansatz~\eqref{eq:ansatz} in which all oscillators are locked to the pacemaker, sufficiently strong coupling is required for the emergence of TL.
In the weakly locked regime, the oscillators do not closely follow the pacemaker.
This makes the value of $\eps_i$ large in Eq~(\ref{eq:ansatz}) and may consequently violate the assumption of the theory.

\subsection{Non-self-oscillatory population}
\begin{figure}[tb]
 \centering
        \includegraphics[width=130mm]{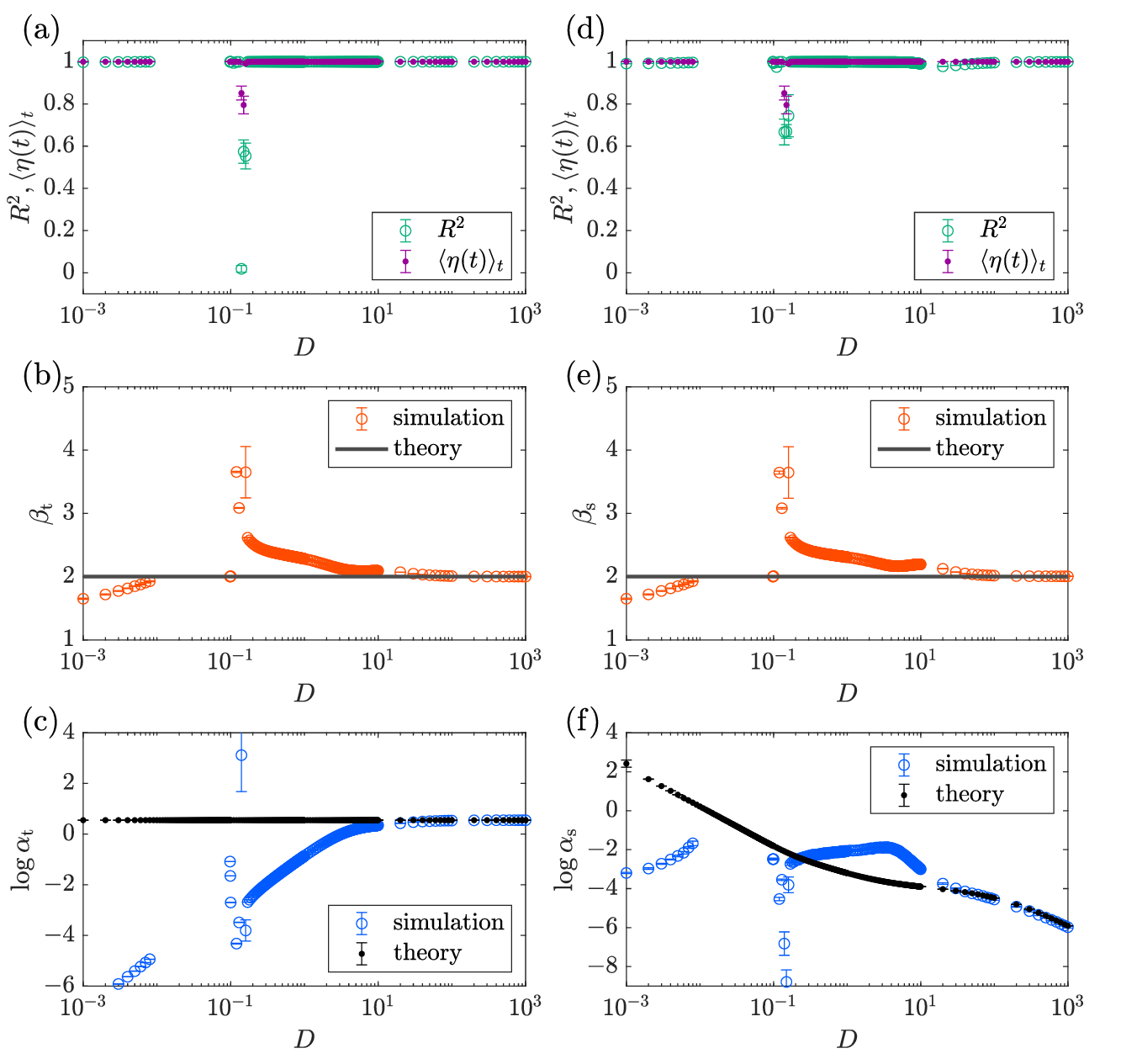}
 \caption{
 Dependence of TL parameters and synchronization degree on $D$ in a population of non-self-oscillatory R\"ossler systems coupled to a pacemaker oscillator. $N=100, \omega_{\mathrm p}=5.5$, $a = 0.1$, $b=0.1$, $c=0.7$, and $\tau = -5.26\times 10^{-4}$. $\omega_i$ is randomly drawn from a uniform distribution between $0.4$ and $0.6$, which makes each element be a damped oscillator. (a)The coefficient of determination, $R^2$, for temporal TL and the time-averaged Kuramoto order parameter $\langle \eta(t) \rangle_t$. (b) Exponents of temporal TL. (c) Intercepts of temporal TL. (d) The coefficient of determination, $R^2$, for spatial TL and the time-averaged Kuramoto order parameter $\langle \eta(t) \rangle_t$. (e) Exponents of spatial TL. (f) Intercepts of spatial TL.}
\label{fig:pacemaker_Rossler_TL_vs_D_with_non_self_oscillatory_population}
\end{figure}

Thus far, we have presented results for a population of limit-cycle oscillators driven by a pacemaker. However, our theoretical investigation is based on the ansatz~(\ref{eq:ansatz}), in which each element is closely entrained by the pacemaker, and this condition does not necessarily require the driven elements to exhibit autonomous oscillations.
To verify this, numerical simulations are conducted on a population of damped oscillators driven by a pacemaker. Figure~\ref{fig:pacemaker_Rossler_TL_vs_D_with_non_self_oscillatory_population} shows the results for both temporal [Figs.~\ref{fig:pacemaker_Rossler_TL_vs_D_with_non_self_oscillatory_population}(a)--(c)] and spatial [Figs.~\ref{fig:pacemaker_Rossler_TL_vs_D_with_non_self_oscillatory_population}(d)--(f)] TLs.
It is confirmed that both temporal and spatial TLs emerge with sufficiently strong coupling.
Note that, in this system as well, the synchronization of the driven elements is achieved even with weak coupling ($D\simeq 0.005, \langle \eta(t) \rangle_t \simeq 1$).
As the coupling constant $D$ increases, synchronization is once weakened for intermediate coupling strengths ($D\simeq 0.15, \langle \eta(t) \rangle_t \simeq 0.8$) and recovers with sufficiently strong coupling ($D\simeq 100, \langle \eta(t) \rangle_t \simeq 1$), and both temporal and spatial TLs emerge in the strongly locked regime. 

\section{Discussion}
In this study, we have investigated a population of R\"ossler oscillators driven by a pacemaker and demonstrated that WP and TL with an exponent 2 can emerge in the system.
We have developed a theory to account for the emergence of TL based on the ansatz in which each driven element is closely entrained by the pacemaker.
This theory indicates that TL can emerge when the intrinsic dynamics of the pacemaker is sufficiently fast and the entrainment of the driven elements is strong.
This theory also suggests that the emergence of TL does not necessarily require fast intrinsic dynamics or autonomous oscillations of the driven elements.
We have confirmed these predictions numerically. These results extend a previous study, which suggested that fast intrinsic oscillations are required for individual elements in coupled systems~\cite{Mitsui_Kori_2025_PRL}, and indicate that TL can emerge in a broad class of synchronized populations.

We have also investigated the relationship between synchronization of the driven elements and the emergence of TL. Through numerical analysis of the phase-locked dynamics of the driven elements, we have found that each element exhibits bistability between two distinct states: a ``weakly locked'' state, in which the element is frequency-synchronized while maintaining small-amplitude oscillations that is far from the pacemaker state, and a ``strongly locked'' state, in which the element is closely entrained to the pacemaker.
We have confirmed that TL emerges when all elements are in the strongly locked regime, consistent with the proposed theory. In this particular system, the emergence of TL is associated with the discontinuous transition in which low-amplitude oscillations vanish abrubtly. Particularly, TL emerges when frequency synchronization involves WP. Synchronization phenomena in which time series become proportional to each other have been previously reported \cite{Mainieri_Rehacek_1999_PRL, generalized_sync_lineartrans, Kano_Umeno_2022_Chaos}. Together with these earlier studies and  previous results on WP \cite{Mitsui_Kori_2025_PRL, Mitsui_Kori_arXiv2025}, the findings of the present study support the universality of this type of synchronization.

In the Kuramoto model, increasing the coupling strength induces a synchronization transition in which the order parameter rises from $0$, a hallmark of the onset of phase synchronization. A similar sharp increase in the order parameter is also observed in our system when the individual elements are achieving weakly locked regime. Importantly, however, the emergence of WP requires substantially stronger coupling. As the coupling strength increases, the synchronization of the entire system once weakens owing to the coexistence of driven elements that are either weakly or strongly locked. When the coupling becomes sufficiently strong, all the elements become strongly locked, resulting in the emergence of WP. The onset of phase synchronization is characterized by phase dynamics. In contrast, WP is characterized by both phase and amplitude, and therefore cannot be captured by phase dynamics alone.

The Kuramoto model considers the coupling strength as a small parameter and successfully establishes a general theoretical framework in the weak-coupling regime. A vast body of work has been conducted using this model to investigate the synchronization phenomena \cite{KuramotoModel_review_2005,KuramotoModel_review_2016}. In contrast, previous studies \cite{Mitsui_Kori_2025_PRL, Mitsui_Kori_arXiv2025} and the present study treat the inverse of the coupling strength as a small parameter and aim to develop a theory of synchronization phenomena in the strong-coupling regime. Recent advances in reduction methods for strongly perturbed or strongly coupled systems \cite{Kurebayashi_etal_PRL2013, reduction_Wilson_PRE2020, Kurebayashi_etal_PRR2022} have enabled phase oscillator models to incorporate amplitude effects into their phase response. These reduction methods may provide simpler dynamical systems capable of explaining collective phenomena in which amplitude degrees of freedom play an essential role---such as WP and TL---in strongly coupled oscillator populations.

\section{Appendix}
\subsection{Comparison of order parameters}
\begin{figure}[tb]
 \centering
        \includegraphics[width=90mm]{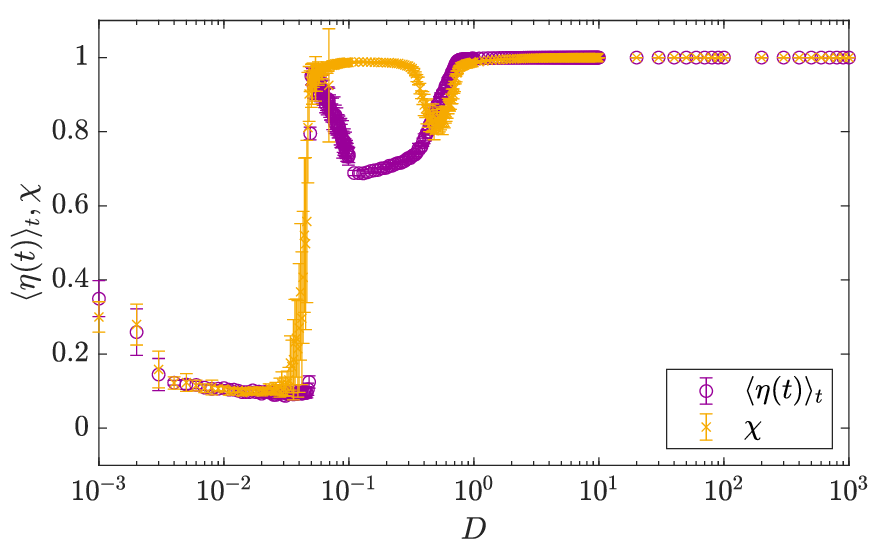}
 \caption{Dependence of the two order parameters, $\langle \eta(t) \rangle_t$ and $\chi$, on $D$ in the pacemaker-driven R\"ossler system. $N=100, \omega_{\mathrm p}=5.5, a = 0.1$, $b=0.1$, and $c=0.7$. $\omega_i$ is randomly drawn from a uniform distribution between $5.4$ and $5.6$.}
\label{fig:eta_chi}
\end{figure}

Figure~\ref{fig:eta_chi} shows the behavior of the two order parameters, $\langle \eta(t) \rangle_t$ and $\chi$, against the coupling strength $D$. Both of them increase sharply at a certain coupling strength and take values close to 1. As the coupling strength is increased further, they decrease once, then begin to increase again, and eventually approach 1.

\subsection{Computing $\tau$}
\begin{figure}[tb]
 \centering
        \includegraphics[width=140mm]{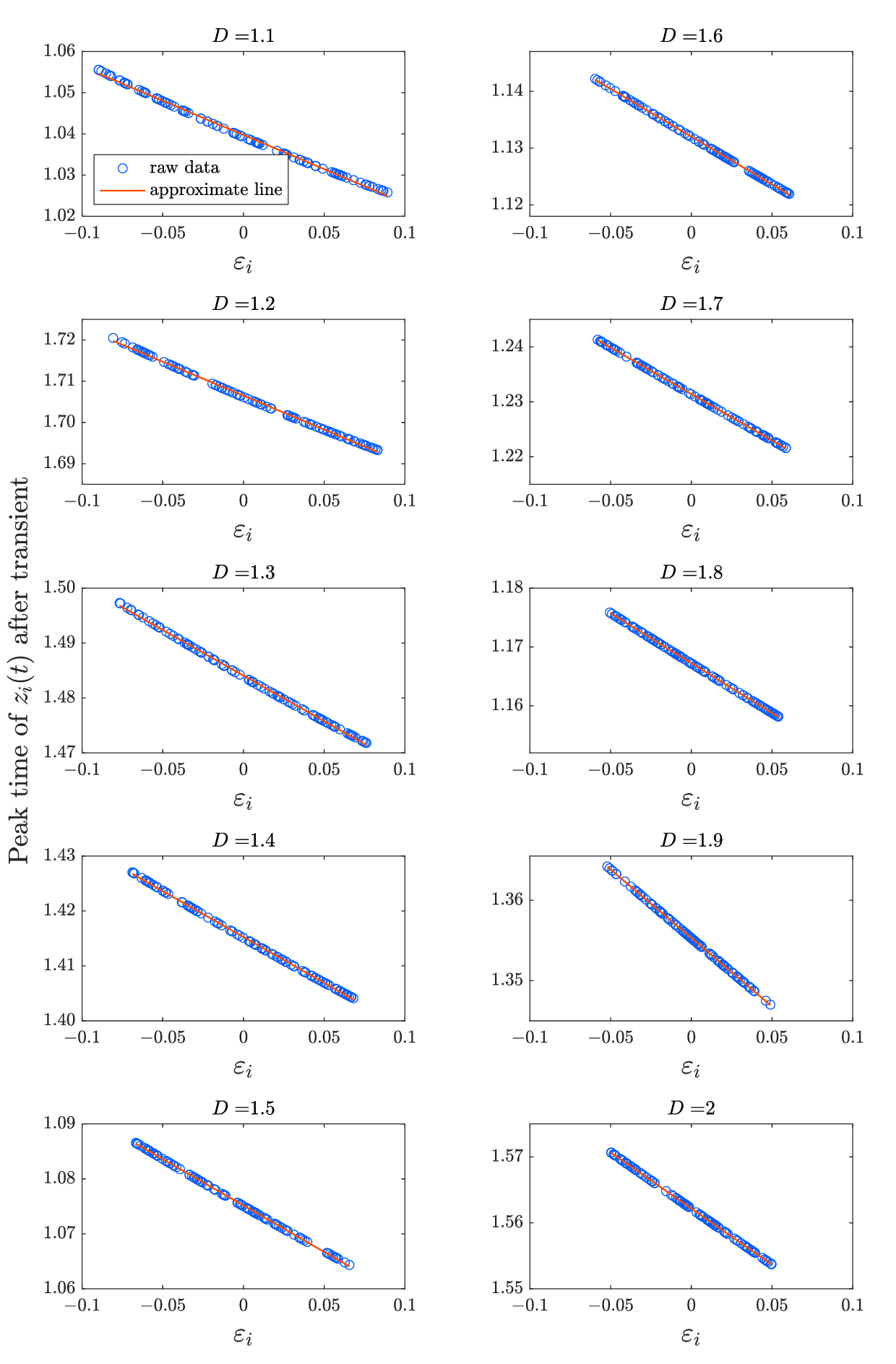}
 \caption{Examples for computing $\tau$ in the pacemaker-driven R\"ossler system. The peak times of each $z_i(t)$ are plotted against $\eps_i$. Data after $t=3000$ are used. $N=100, \omega_{\rm p}=5.5, a = 0.1$, $b=0.1$, and $c=0.7$. $\omega_i$ is randomly drawn from a uniform distribution between $5.4$ and $5.6$. The blue dots represent raw data, and the red lines represent linear fitting lines.}
\label{fig:phase_lag}
\end{figure}

We compute $\tau$ by plotting the peak times of each time series against $\eps_i$ and performing a linear fitting (Fig.~\ref{fig:phase_lag}). We repeat this procedure for ten different coupling strengths and take the average value as $\tau$. In this analysis, we are only concerned with the slopes; the intercepts has no meaning.

%
%

\ack{Y. M. thanks Hiroshi Ito for helpful discussions.
Y. M. acknowledges the Education and Research Center for Mathematical and Data Science, Kyushu University and JSPS KAKENHI 26KJ0246 for financial support. S. H. acknowledges JSPS KAKENHI 25K15264 for financial support. H. K. acknowledges JSPS KAKENHI 25K15258 for financial support.}


\roles{Yuzuru Mitsui https://orcid.org/0009-0005-4019-5468\\
\noindent Conceptualization, Formal analysis, Methodology, Visualization, Writing – original draft (lead)\\
\noindent Shigefumi Hata https://orcid.org/0000-0002-6271-6982\\
\noindent Writing – original draft (supporting), Writing – review \& editing (lead)\\
\noindent Hiroshi Kori https://orcid.org/0000-0003-2899-7896\\
\noindent Supervision, Writing – review \& editing (supporting)}


\data{The data that support the findings of this study are available
from the corresponding author upon reasonable request.}


\clearpage
\bibliographystyle{iopart-num} 
\bibliography{ref}

\end{document}